\documentclass{article}
\usepackage{spconf,amsmath,graphicx}

\usepackage{enumitem}
\setlist{nosep, leftmargin=14pt}

\usepackage{cite}
\usepackage{amsmath,amssymb,amsfonts}
\usepackage{algorithmic}
\usepackage{graphicx}
\usepackage{textcomp}
\usepackage{xcolor}
\def\BibTeX{{\rm B\kern-.05em{\sc i\kern-.025em b}\kern-.08em
    T\kern-.1667em\lower.7ex\hbox{E}\kern-.125emX}}

\usepackage{cleveref}
 \usepackage{url}
\usepackage{tikz}
\usetikzlibrary{external}
\usetikzlibrary{calc}
\usetikzlibrary{shapes.arrows, patterns, arrows.meta}
\usetikzlibrary{positioning,backgrounds}
\usepackage{url}

\usepackage{pgfplots}
\usepgfplotslibrary{groupplots,colormaps}

\usepackage{pgfplotstable}
\pgfplotsset{compat=1.16}

\pgfplotsset{
    colormap={LocalControlYeswhite}{
        rgb255=(112,128,144)
        rgb255=(112,128,144)
    },
}
\pgfplotsset{
colormap={hotred}{color(0cm)=(orange); color(1cm)=(orange); color(2cm)=(orange); color(3cm)=(orange)}
}

\definecolor{patientAB}{HTML}{fad50b}
\definecolor{patientCD}{HTML}{02fa57}
\definecolor{patientEF}{HTML}{778bfa}
\definecolor{patientGH}{HTML}{eb6134}

\DeclareRobustCommand\circled[2]{
    \tikz[baseline=0pt]{
        \node[anchor=base, shape=circle,inner sep=0.0pt, minimum size=2.5mm, fill=#1, text=white] (char) {\strut #2};
    }
}

\newlength{\px}
\DeclareMathOperator{\atantwo}{arctan2}

\name{Frederic Madesta$^{\dagger \ddagger}$ \qquad Lukas Wimmert$^{\dagger \ddagger}$ \qquad Tobias Gauer$^{\star}$ \qquad Ren\'e Werner$^{\dagger \ddagger}\sthanks{Shared last authorship.\vspace*{0.3cm}\newline This work has been submitted to the IEEE for possible publication. Copyright may be transferred without notice, after which this version may no longer be accessible.}$ \qquad Thilo Sentker$^{\dagger \ddagger *}$}

\address{$^{\dagger}$ Institute for Applied Medical Informatics, University Medical Center Hamburg-Eppendorf\\
         $^{\ddagger}$ Institute of Computational Neuroscience, University Medical Center Hamburg-Eppendorf\\
         $^{\star}$ Department of Radiotherapy and Radiation Oncology, University Medical Center Hamburg-Eppendorf \vspace{-0.1cm}
         }
\title{Oriented Histogram-Based Vector Field Embedding for Characterizing 4D CT Data Sets in Radiotherapy}

\begin{document}


\maketitle

\begin{abstract}
In lung radiotherapy, the primary objective is to optimize treatment outcomes by minimizing exposure to healthy tissues while delivering the prescribed dose to the target volume. The challenge lies in accounting for lung tissue motion due to breathing, which impacts precise treatment alignment. To address this, the paper proposes a prospective approach that relies solely on pre-treatment information, such as planning CT scans and derived data like vector fields from deformable image registration. This data is compared to analogous patient data to tailor treatment strategies, i.e., to be able to review treatment parameters and success for similar patients.

To allow for such a comparison, an embedding and clustering strategy of prospective patient data is needed. Therefore, the main focus of this study lies on reducing the dimensionality of deformable registration-based vector fields by employing a voxel-wise spherical coordinate transformation and a low-dimensional 2D oriented histogram representation. Afterwards, a fully unsupervised UMAP embedding of the encoded vector fields (i.e., patient-specific motion information) becomes applicable. The functionality of the proposed method is demonstrated with 71 in-house acquired 4D CT data sets and 33 external 4D CT data sets. A comprehensive analysis of the patient clusters is conducted, focusing on the similarity of breathing patterns of clustered patients. 
The proposed general approach of reducing the dimensionality of registration vector fields by encoding the inherent information into oriented histograms is, however, applicable to other tasks. 
\end{abstract}


\section{Introduction}
In current lung cancer radiotherapy (RT), advanced imaging techniques play a pivotal role in the treatment process to optimize patient outcomes \cite{schmitt:2020}. The main objective remains to minimize the exposure of healthy tissue while ensuring that the planned dose is delivered to the target volume. Consideration of the physiological motion of lung tissues due to breathing is a crucial factor during treatment planning and dose delivery, as it has a direct impact on treatment success. Established approaches utilize time-resolved computed tomography (4D CT) to account for motion during RT. Such methods involve irradiating the entire volume within which the target moves, gated dose delivery, or tumor tracking. 

In clinical practice, the obtained 4D imaging data usually serves exclusively for immediate and patient-specific treatment purposes, with only the direct image data being used. Potentially relevant derived data like motion statistics are not part of established current workflows. Motion information such as vector fields computed by deformable image registration (DIR) and representing motion between distinct 4D CT phases, for instance, can be used for retrospective dose simulation and correlation of respective dose distributions to clinical endpoints \cite{sothmann:2017}. However, at the moment, such motion fields and related data are mainly considered to be of academic interest rather than providing valuable information. 

Thus, from our perspective, the potential of the analysis of motion field statistics at a patient and a population level is not fully realized in clinical practice. We think that a prospective analysis of lung motion statistics, i.e., before generating the treatment plan, could offer relevant insight into potential issues during treatment.  

In the present study, the aim is to take this a step further. More specifically, the idea is to use 4D CT-derived motion data to compare it to previously acquired, closely analogous patient data, e.g., to employ a similar treatment approach if it was successful for these patients; otherwise, the treatment would have to be altered. In an ideal scenario, this might even enable the motion-based prediction of treatment success probability. In turn, this could facilitate an adjustment of the treatment plan or a more careful evaluation, particularly in cases where the likelihood of an unsuccessful treatment is anticipated.

However, analyzing and clustering high-dimensional motion data in a meaningful way remains a complex task, and applying some sort of dimensionality reduction technique beforehand is mandatory, with the challenge that the reduction of high-dimensional data to a lower-dimensional representation carries an inherent risk of losing critical information during the process. 
Here, we introduce an oriented histogram-based approach for dimensionality reduction of motion fields that is designed to mitigate this risk and retain relevant motion information. As far as we are aware, such a dimensionality reduction for high dimensional vector field data has not been explored in the existing literature, highlighting the unique perspective offered by our proposed approach.
To showcase the potential of our approach, we employ the encoded data for embedding the vector fields in a 2D manifold and thereby characterizing 4D CT datasets. Taking up the aforementioned challenges,
\begin{itemize}
    \item[\textbf{C1}] we introduce a straightforward method of low computational complexity (code available upon acceptance at \newline \url{github.com/IPMI-ICNS-UKE/vf-clustering}) for reducing the dimensionality of DIR-based vector fields, with the reduction being achieved by application of a voxel-wise spherical coordinate transform, followed by the computation of a weighted 2D oriented histogram,
    \item[\textbf{C2}] demonstrate the feasibility and interpretability of the aforementioned approach to encode the vector field information extracted from two distinct 4D CT data sets,
    \item[\textbf{C3}] embed the encoded patient-specific internal motion information stored in the oriented histograms using an autoencoder and Uniform Manifold Approximation and Projection (UMAP),
    \item[\textbf{C4}] and perform a comprehensive analysis of the patient clusters generated in the UMAP (regarding similarity of breathing patterns) for both an in-house as well as an external 4D CT data set.
\end{itemize}
To the best of our knowledge, all aspects C1-C4 are novel contributions in the given application context. 

\section{Methods}
\subsection{Datasets}
\label{sec:data}
We utilized anonymized 4D CT data from 71 lung cancer patients (in-house data set collected between 2015 and 2020). 4D CT image reconstruction was performed using phase-based (34 cases) and amplitude-based (37 cases) binning techniques, with the resulting 4D CT dataset comprising $n_{\text{ph}}=10$ 3D CT phase images.
Additionally, we employed a subset of the publicly available 4D-Lung dataset \cite{balik:2013,hugo:2016,hugo:2017} comprising 33 artifact-free 10-phase 4D CT data sets from 8 lung cancer patients, each containing 2 to 6 repeat 4D CT images. 

\subsection{Deformable image registration}
\begin{figure*}[pth]
    \centering
    \input{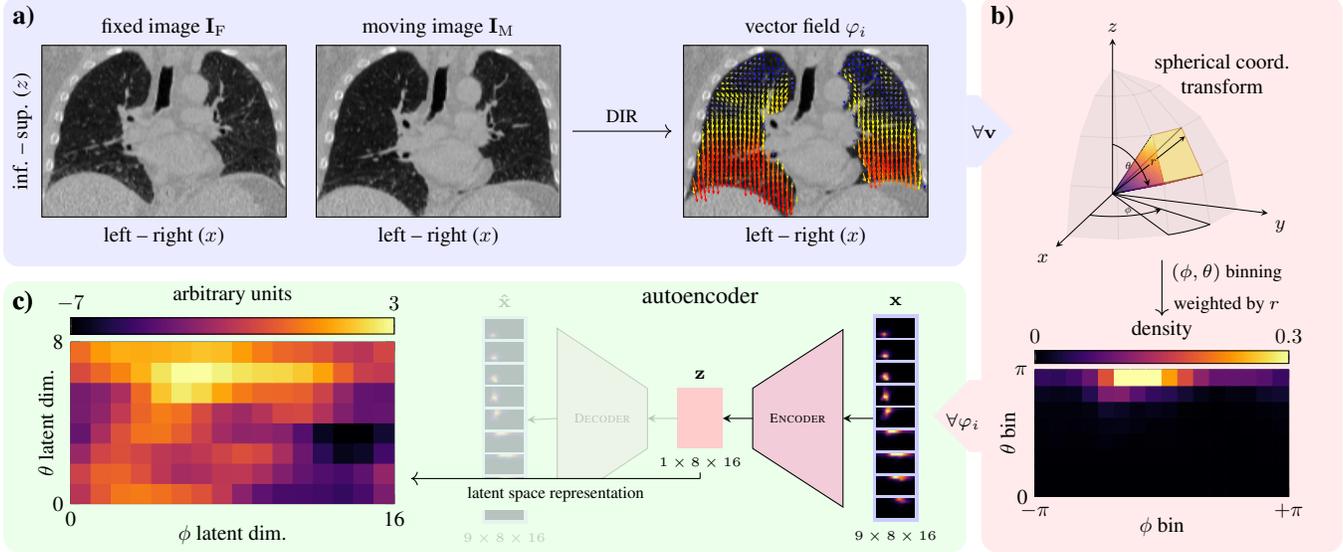}
    \caption{Proposed workflow for reducing the dimensionality of 4D vector field data of a 4D CT patient data set. Block a) DIR of fixed and moving images yields high-dimensional vector field data. Block b) By applying a spherical coordinate transformation to each vector $\mathbf{v}$ in the field, an oriented histogram can be extracted through $r$-weighted $(\phi, \theta)$ binning. Performing this process for all phases of a 4D CT data set results in 9 oriented histograms. In block c), the data is further encoded by extracting the corresponding latent space representation using a lightweight autoencoder.}
    \vspace{-0.25cm}
    \label{fig:flow}
\end{figure*}

We employed a GPU implementation of the deformable demons-based Variational Registration algorithm known for its efficacy in lung CT registration \cite{werner:2014}. In the context of 4D CT registration, one of the phase images was selected as the reference, i.e., the fixed image, ($\mathbf{I}_j \equiv \mathbf{I}_\text{F}$), while the other phase images served as the moving images ($\mathbf{I}_{i\neq j} \equiv \mathbf{I}_\text{M}$), cf. \cref{fig:flow} block a). These moving images were sequentially registered to $\mathbf{I}_\text{F}$ by minimizing the objective function $\varphi_i = \arg\min_{\varphi_i^\ast\in\mathcal{C}^2[\Omega]} \mathcal{J}\left[\mathbf{I}_\text{F},\mathbf{I}_\text{M};\varphi_i^\ast\right]$, yielding the sought transformations $\varphi_{i \in \{1, \ldots, n_{\text{ph}}\}\setminus\{j\}}$ for all $n_{\text{ph}}-1$ image pairs.

The computation of $\varphi$ involved the application of active demon forces, leading to the iterative update step $\Delta\varphi_{i+1} = \mathcal{R}\left(\Delta\varphi_i - \tau \mathbf{F}\right)$, with \(\tau \in \mathbb{R}^+\) representing the time step (in this case, $\tau = 2.25$). Force computation was defined as 
\begin{equation*}
\mathbf{F}(\mathbf{I}_\mathrm{F}, \mathbf{I}_\mathrm{M}, \varphi_i) = \frac{\mathbf{I}_\mathrm{F}-\mathbf{I}_\mathrm{M}\circ\varphi_i}
{\|\nabla\mathbf{I}_\mathrm{M}\circ\varphi_i\|^2+\alpha(\mathbf{I}_\mathrm{F}-\mathbf{I}_\mathrm{M}\circ\varphi_i)^2}
\nabla\mathbf{I}_\mathrm{M}\circ\varphi_i
\end{equation*}
where $\alpha\in\mathbb{R}^+$ denotes the inverse of the mean squared image spacing and $\mathcal{R}$ represents Gaussian regularization with $\sigma=(1.25, 1.25, 1.25)$ voxels. Detailed discussions on the influence of the hyperparameters $\tau$ and $\sigma$ on the registration results can be found in \cite{werner:2014}.

For the sequential registration of a single 4D CT data set, we obtained 9 vector fields, each representing the registration of one reference phase to all the remaining phases. One vector field has a size of $[I_x, I_y, I_z, 3]$, corresponding to the input image size with stored 3D vector information in each voxel.
\subsection{Oriented histogram of 3D vector fields}
The comparison of inter- and intra-patient 4D CT data requires the alignment of the images itself or the vector fields, as the patient orientation usually differs between the scans.
In this work, we opted for changing the vector field basis to the lung basis, which is defined by the principal axes of orientation of the lung mask.
Let $\mathbf{C}\in\mathbb{R}^{3\times n_\mathrm{m}}$ be the matrix of the $n_\mathrm{m}$ mean-centered coordinates of non-zero lung mask voxels.
We then computed the three eigenvectors of the covariance matrix $\mathbf{\Sigma}_{CC}\in\mathbb{R}^{3\times 3}$.
Given the 4D CT base $(\mathbf{e}_x, \mathbf{e}_y, \mathbf{e}_z$), we subsequently chosen the nearest right-handed orthonormal basis constructed from the eigenvectors as the lung basis
\begin{equation}
    \mathbf{B}_\mathrm{lung}=(\mathbf{e}_1, \mathbf{e}_2, \mathbf{e}_3)^\top\in\mathbb{R}^{3\times 3}.
\end{equation}
The vector fields $ \mathbf{V}\in\mathbb{R}^{3\times I_x\cdot I_y\cdot I_z}$ were then transformed from the 4D CT basis to the lung basis by
\begin{equation}
    \mathbf{V}_\mathrm{lung} = \mathbf{B}_\mathrm{lung}^{-1} \mathbf{V}.
\end{equation}

Clustering the high-dimensional vector fields resulting from the applied registration approach proved to be a challenging task. To address this, we opted for dimensionality reduction through the use of histograms of oriented optical flow \cite{chaudhry:2009} (hereinafter referred to as oriented histograms), which we extended to 3D vector fields.
Here, we encoded the direction $(x, y, z)$ of each vector in the vector field $\mathbf{V}_\mathrm{lung}$ by applying a spherical transformation, i.e., 
\begin{equation*}
   r=\sqrt{x^2 + y^2 + z^2}, \quad \theta = \arccos\left(\frac{z}{r}\right), \quad \phi = \atantwo\left(\frac{y}{x}\right)
\end{equation*} 
and subsequently tallied the occurrences of vectors inside the lung masks pointing in a similar direction defined by bins of size $\frac{\pi}{8}$ for $\theta\in[0,\pi]$ and $\phi\in (-\pi,\pi]$. To reduce the noise introduced by vectors with negligible magnitude, each histogram bin entry was weighted by its corresponding radius $r\in \mathbb{R}^+$. As a last step, the oriented histogram is normalized by the number of lung voxels taken into account. This process effectively reduced a vector field of size $[I_x, I_y, I_z, 3]$ to a compact representation of size $8\times16$ (as indicated in \cref{fig:flow} block b)), albeit at the cost of losing specific localization information while retaining essential insights into the distribution of lung motion directions and magnitudes. The encoding of vector fields for all 10 phases of a 4D CT dataset leads to the generation of 9 oriented histograms. Due to the sparse nature of the oriented histograms, a lightweight convolutional autoencoder was trained and applied to extract even more compressed but now denser latent space representations (cf. \cref{fig:flow} block c)). The autoencoder is solely trained with an $L^2$ reconstruction loss and thus is a fully unsupervised feature extractor.
It has been shown that this autoencoding step prior to local manifold learning is an effective approach to discover higher quality clusters \cite{mcconville:2021}.
\subsection{UMAP embedding}
We utilized UMAP \cite{mcinnes:2018}, a non-linear dimensionality reduction technique, to analyze and characterize the encoded vector fields (i.e., the latent space representations of the oriented histograms) after embedding them in a 2D manifold. 
Here, we used the following key parameters: 5 as size of local neighborhood, cosine distance as metric, and DensMAP \cite{narayan:2021} as local density regularization.
\begin{figure*}
    \centering
        \input{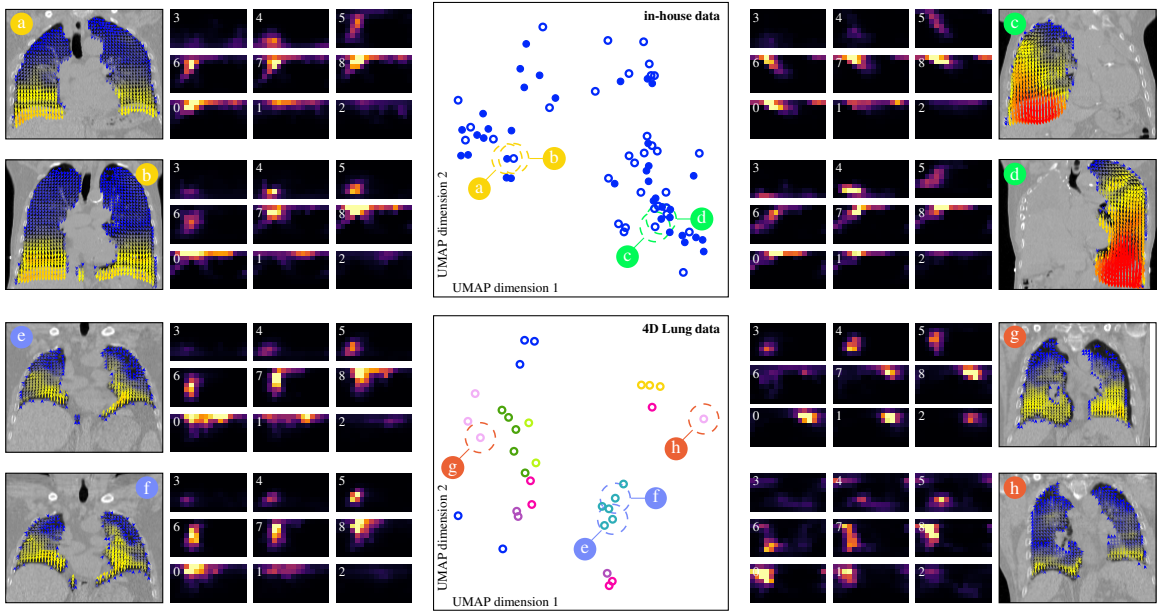}
        \vspace{-0.5cm}
        \caption{UMAP embedding of the 71 in-house (top) and 33 4D Lung (bottom, mark color coding: patient) encoded vector field sequences after embedding them in a 2D manifold (middle, scatter plots) and illustration of selected data points by corresponding image data and all 9 oriented histograms (subfigures a--d [in-house data] and e--h [4D Lung data], white number: moving phase). Amplitude- and phase-based reconstructed 4D CT datasets are marked with $\bullet$ and $\circ$, respectively. 
        Subplots \circled{patientAB}{a} and \circled{patientAB}{b} show two patients whose vector fields and corresponding oriented histograms remarkably resemble each other, despite different lung volumes.
        Subplots \circled{patientCD}{c} and \circled{patientCD}{d} illustrate two patients with only the right and left lung, respectively, yet having similar vector fields characteristics.
        \textcolor{black}{Subplots} \circled{patientEF}{e} and \circled{patientEF}{f} depict repeat 4D CTs of one patient with high similarity. 
        \textcolor{black}{Subplots} \circled{patientGH}{g} and \circled{patientGH}{h} illustrate repeat 4D CTs of one patient that show large deviations in both the motion vector fields and the oriented histograms.
        }
        \vspace{-0.25cm}
    \label{fig:results}
\end{figure*}
\section{Results}
All patient datasets underwent the registration and dimensionality reduction process outlined in \cref{fig:flow}. We selected the mid ventilation phase as the fixed image, i.e., $\mathbf{I}_\mathrm{F}=\mathbf{I}_{j=3}$, to obtain both positive and negative $z$-displacements, thus utilizing the full $\theta$-range of the oriented histograms. Afterward, the encoded data was subjected to analysis within the 2D UMAP manifold, with the focus lying on clustering with respect to breathing pattern similarity. 

As shown in \cref{fig:results}, patients with vector fields with similar characteristics, including but not limited to overall smoothness, linearity and angular distribution tend to form a cluster. This becomes especially apparent for the repeat 4D CT data of the 4D Lung data set, where most of the repeat 4D CT data clusters, and, if not, large motion vector field deviations between data points are present.
Clustering based on even more apparent features like phase- vs. amplitude-based binning during image reconstruction or large vs. small motion amplitudes were not observed. The latter may be a direct consequence of the employed cosine distance.
The UMAP embedding yielded several smaller clusters to be further investigated. In addition, the proposed oriented histogram and its embeddings seem to be robust against anatomical outliers, as demonstrated in \cref{fig:results}c/d for two post-pneumonectomy patients.
\section{Discussion and Conclusions}
The presented embedding of vector fields computed by DIR of 4D CT data by applying oriented histograms to reduce dimensionality shows that the methodological approach is feasible to allow for analyses of clusters in the UMAP space. The clusters exhibited similar breathing patterns of the patients, illustrating that the embedding appears to be meaningful, which is further demonstrated by the patient-wise clustering of the repeat 4D CT data of the 4D Lung data set. 

As demonstrated in \cref{fig:results}, the oriented histogram proves robust against anatomical scenarios where a principal component analysis of the vector fields would fail in quantifying breathing pattern similarity. It is subject to future research to evaluate the effect of the angular resolution used for $(\phi, \theta)$ binning. Choosing a smaller bin size than the applied $\frac{\pi}{8}$ would result in more detailed oriented histograms and thus may enable UMAP to incorporate more complex breathing patterns. 

From an application perspective, it should be emphasized that no supervised learning is employed in any step of the presented workflow. 
Clustering seems to be independent of the 4D CT reconstruction approach (amplitude- vs. phase-based). Although the datasets used in this study comprise 71 and 33 data points, which are relatively large for the specific clinical field, their sizes are still limited and a more comprehensive evaluation would require additional data points. However, the present contribution focused on the methodical developments. 

Independent of the specific field of application, the proposed method of computational efficient dimensionality reduction for registration vector fields, achieved through encoding the inherent information into oriented histograms, remains a practical and adaptable approach for characterizing motion patterns of patients and even beyond, if it is applied to other registration tasks and anatomies.  

The proposed approach also enables applications that go beyond the pure DIR-based analysis of anatomical changes or physiological processes. During the Learn2Reg MICCAI challenges 2022 \& 2023 on lung CT image registration, we (VROC, one of the top-performing submissions) successfully applied the embedding concept to choose patient-specific optimal DIR-hyperparameters based on the nearest neighbors in the UMAP. The concept is applicable to any DIR algorithm that relies on manually set hyperparameters, and the present publication is the first to explain the underlying methodology.
\section{Acknowledgments}
This work was funded by DFG research grant \mbox{WE 6197/2-2} (project number 390567362). 
\section{Compliance with Ethical Standards}
The study was conducted retrospectively using human subject data. The data evaluation was approved by the local ethics board, and the requirement to obtain written informed consent was waived [WF-82/18] between 2015 and 2020.

\bibliographystyle{IEEEtran.bst}
\bibliography{IEEEabrv, main_isbi}

\begin{thebibliography}{10}
\providecommand{\url}[1]{#1}
\csname url@samestyle\endcsname
\providecommand{\newblock}{\relax}
\providecommand{\bibinfo}[2]{#2}
\providecommand{\BIBentrySTDinterwordspacing}{\spaceskip=0pt\relax}
\providecommand{\BIBentryALTinterwordstretchfactor}{4}
\providecommand{\BIBentryALTinterwordspacing}{\spaceskip=\fontdimen2\font plus
\BIBentryALTinterwordstretchfactor\fontdimen3\font minus
  \fontdimen4\font\relax}
\providecommand{\BIBforeignlanguage}[2]{{%
\expandafter\ifx\csname l@#1\endcsname\relax
\typeout{** WARNING: IEEEtran.bst: No hyphenation pattern has been}%
\typeout{** loaded for the language `#1'. Using the pattern for}%
\typeout{** the default language instead.}%
\else
\language=\csname l@#1\endcsname
\fi
#2}}
\providecommand{\BIBdecl}{\relax}
\BIBdecl

\bibitem{schmitt:2020}
D.~Schmitt, O.~Blanck, T.~Gauer, M.~K. Fix, T.~B. Brunner, J.~Fleckenstein,
  B.~Loutfi-Krauss, P.~Manser, R.~Werner, M.~L. Wilhelm, W.~W. Baus, and
  C.~Moustakis, ``Technological quality requirements for stereotactic
  radiotherapy,'' \emph{Strahlenther Onkol}, vol. 196, no.~5, pp. 421--43,
  2020.

\bibitem{sothmann:2017}
T.~Sothmann, T.~Gauer, M.~Wilms, and R.~Werner, ``Correspondence model-based 4d
  vmat dose simulation for analysis of local metastasis recurrence after
  extracranial sbrt,'' \emph{Phys Med Biol}, vol.~62, no.~23, p. 9001, 2017.

\bibitem{balik:2013}
S.~Balik, E.~Weiss, N.~Jan, N.~Roman, W.~C. Sleeman, M.~Fatyga, G.~E.
  Christensen, C.~Zhang, M.~J. Murphy, J.~Lu \emph{et~al.}, ``Evaluation of
  4-dimensional computed tomography to 4-dimensional cone-beam computed
  tomography deformable image registration for lung cancer adaptive radiation
  therapy,'' \emph{Int J Radiat Oncol Biol Phys}, vol.~86, no.~2, pp. 372--379,
  2013.

\bibitem{hugo:2016}
G.~D. Hugo, E.~Weiss, W.~C. Sleeman, S.~Balik, P.~J. Keall, J.~Lu, and J.~F.
  Williamson, ``Data from {4D} lung imaging of {NSCLC} patients,'' \emph{Cancer
  Imaging Arch}, vol.~10, p.~K9, 2016.

\bibitem{hugo:2017}
------, ``A longitudinal four-dimensional computed tomography and cone beam
  computed tomography dataset for image-guided radiation therapy research in
  lung cancer,'' \emph{Med Phys}, vol.~44, no.~2, pp. 762--771, 2017.

\bibitem{werner:2014}
R.~Werner, A.~Schmidt-Richberg, H.~Handels, and J.~Ehrhardt, ``Estimation of
  lung motion fields in {4D} {CT} data by variational non-linear
  intensity-based registration: {A} comparison and evaluation study.''
  \emph{Phys Med Biol}, vol.~59, no.~15, pp. 4247--60, 2014.

\bibitem{chaudhry:2009}
R.~Chaudhry, A.~Ravichandran, G.~Hager, and R.~Vidal, ``Histograms of oriented
  optical flow and binet-cauchy kernels on nonlinear dynamical systems for the
  recognition of human actions,'' in \emph{2009 IEEE conference on computer
  vision and pattern recognition}.\hskip 1em plus 0.5em minus 0.4em\relax IEEE,
  2009, pp. 1932--1939.

\bibitem{mcconville:2021}
R.~McConville, R.~Santos-Rodriguez, R.~J. Piechocki, and I.~Craddock, ``N2d:
  (not too) deep clustering via clustering the local manifold of an autoencoded
  embedding,'' in \emph{2020 25th International Conference on Pattern
  Recognition (ICPR)}.\hskip 1em plus 0.5em minus 0.4em\relax IEEE, Jan. 2021.

\bibitem{mcinnes:2018}
L.~McInnes, J.~Healy, and J.~Melville, ``Umap: Uniform manifold approximation
  and projection for dimension reduction,'' \emph{arXiv preprint
  arXiv:1802.03426}, 2018.

\bibitem{narayan:2021}
A.~Narayan, B.~Berger, and H.~Cho, ``Assessing single-cell transcriptomic
  variability through density-preserving data visualization,'' \emph{Nat
  Biotechnol}, vol.~39, no.~6, pp. 765--774, 2021.

\end{thebibliography}

\end{document}